# UPGRADE TO FIXED AND TRANSLATING SCINTILLATION-BASED LOSS DETECTOR SYSTEM IN THE FERMILAB DRIFT TUBE LINAC*

E. V. Chen[1,†], A. L. Saewert[1], R. V. Sharankova[1]
[1]Fermi National Accelerator Laboratory, Batavia, IL, 60510, USA

*Abstract*

The closed-off structure of the Fermilab Drift Tube Linac precludes a robust array of instrumentation from directly monitoring the H⁻ beam that is accelerated from 750 keV to 116 MeV. To improve beam tuning and operational assessment of Drift Tube Linac performance, scintillator-based loss monitors were previously installed along the exterior of the first two accelerating cavities to assess low energy beam losses. Here we present a recent upgrade to the loss monitor system, including significant improvements in analog signal processing to address baseline-interfering noise; digitization of the signals to enable regular operational use and tuning; and a new remote operation upgrade of the translating loss monitor with precise positioning of the loss monitor along its nine-foot track. Data from the fixed and translating detectors collected under varying beam conditions validate the utility of the upgrade.

## INTRODUCTION

The Fermilab Drift Tube Linac (DTL) accelerates H⁻ ions from 750 keV to 116 MeV using five 201 MHz cavities. The DTL has a ~92% transmission efficiency, making it the section of the linac with the highest beam loss. Given the lack of diagnostics within the cavities and the known misalignments of the drift tubes, especially in the vertical direction [1], it is difficult to tune the DTL effectively.

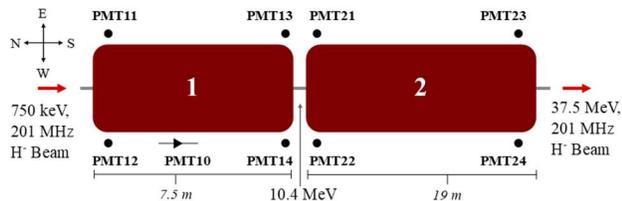

Figure 1: Schematic of the first two DTL cavities and the arrangement of scintillator detectors around the cavities, labelled PMTmn, where m refers to the cavity number and n refers to the position. Circle markers and triangle markers indicate stationary and translating detectors, respectively.

Previously, we demonstrated the ability of scintillator-based detectors to quantify products of beam loss when installed outside of the DTL cavities [2], using a translating detector (PMT10) and two stationary detectors (PMT23, PMT24). The detectors are composed of a scintillator (EJ-208 or NE102), light guide, and photomultiplier tube. At the first two DTL cavities (750 keV to 37.5 MeV), we expect the scintillators to detect x-rays and gamma rays generated from beam loss and RF cavity emissions.

This work describes an upgrade to the scintillator-based detector system along the first two DTL cavities, bringing it to an operational-ready state for use in quadrupole magnet tuning and DTL alignment, and an initial investigation in beam loss patterns.

## DETECTOR UPGRADES

The upgrade involves the installation of six additional stationary detectors (PMT11, PMT12, PMT13, PMT14, PMT21, and PMT22) along the sides of the first two DTL cavities (Fig. 1); digitization of the detector signals for integration with Fermilab's accelerator control system; improved signal processing and amplification of the detector signals; and an upgrade to the translating detector system for remote control and greater detector position precision.

Installation of the new detectors involved tuning the photomultiplier tube supply voltage to yield acceptable signal-to-noise and avoid saturation in the photomultiplier tube. Detectors are not absolutely calibrated and supply voltages range from 600 to 1100 V.

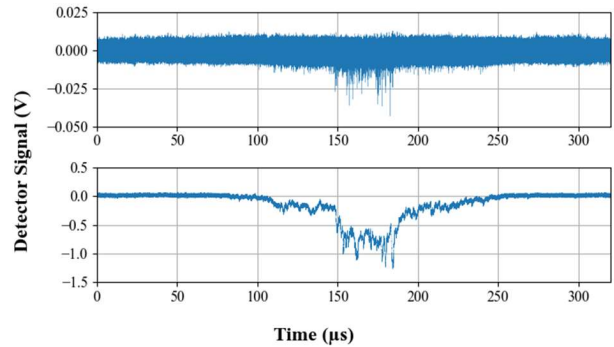

Figure 2: Signal from the PMT14 stationary detector. Raw data (top) and data processed with the amplification and filtering module (bottom) are shown from different 35.64 μs pulses, beginning at ~150 μs. More negative signal corresponds to greater beam loss.

*Signal Amplification, Filtering, and Digitization*

NIM modules were developed to amplify and filter the raw detector signal for digitization (Fig. 2). The modules contain three amplification stages, one non-inverting and two inverting, with an overall gain of 28 μA/V and a bandwidth from DC to 350 kHz. This bandwidth provides integration of the PMT pulses for the 10 MHz, ±5 V digitizers and reduces the 201 MHz Linac gallery RF noise by over 100 dB. A common-mode choke is used at each channel before the amplifier/filter stages to eliminate the ~500 Hz baseline wavelet noise produced by quadrupole power supplies. The choke was made from a compact nanocrystalline core for high permeability, with inductance per square turn

($A_L$) of 88 µH at 10 kHz. With 18 turns, the overall inductance is 28.5 mH at 10 kHz ($A_L = L/N^2$).

The digitizer returns the full 400 µs waveform and also averages over two pre-defined 10 µs windows, one to capture the RF cavity emission and one to capture the beam window. The detector signal corresponding to beam loss is the RF average subtracted from the beam pulse average, as previously described [2]. Data were collected at 3 Hz with a 12 µs pulse length, unless otherwise specified.

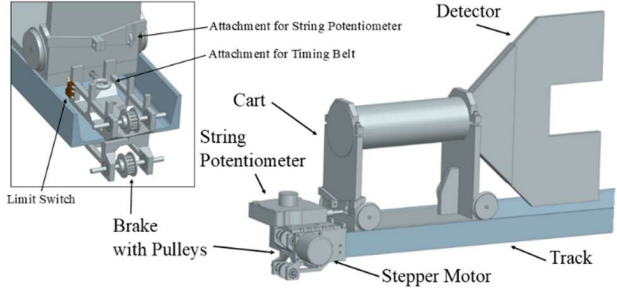

Figure 3: 3D model of the translating detector system on the upstream end of the track. Timing belt and some bolts/nuts are not shown. Close up shows the assembly with the string potentiometer and stepper motor removed.

*Remote Control of Translating Detector*

The translating detector system received a significant upgrade to enable remote control of the cart system and improve the precision of its position readback. The brakes at each end of the track were redesigned to accommodate toothed pulleys, and a matched timing belt was looped around the track and attached to both ends of the cart (Fig. 3). The effective length of the track reduced slightly with the redesigned brakes – from 2.3 to 2.2 meters – but a string potentiometer was attached to the cart to yield a readback of the current position (±0.01 m, a tenfold improvement on the 10 cm increments previously used [2]). A pulley at the upstream end couples to a stepper motor, allowing the cart to travel the track in ~5500 steps, or 0.4 mm/step.

The cart's upstream position is 2.5 m from the start of the first DTL cavity. Data were collected by running the cart at its maximum velocity, with information collected with both upstream and downstream movement.

## VALIDATION OF UPGRADES

Increased beam loss was introduced by scanning the current in a horizontal dipole magnet, MDQ2H, located at the entrance to the DTL. We expected to observe changes in the loss profiles due to trajectory misalignment in the quadrupoles and scraping at the entrance of the cavity.

The results (Fig. 4) show a visible change in losses at all detectors except the most upstream, PMT11, PMT12, and PMT10 (the translating detector). The change in losses ranged from 0.3 V to 1.9 V, higher for the PMTs located further downstream. Notably, at readings with the highest losses, we observed saturation at the digitizer, where points on the digitized waveform were clamped at -5 V (Fig. 6). Data that saturate in the digitizer underestimate losses. This can be mitigated by reducing the gains on the amplification module to decrease the signal strength entering the digitizer; however, for tuning purposes, we are not interested in the regions with highest loss that correspond to significant scraping of the beam.

To be relevant to tuning, the beam current must be preserved, occurring at MDQ2H currents between -0.5 to 0.5 A. Here, changes in losses are less severe, but still detectable. We observe a monotonic change in losses, contrary to expectations of a loss profile with a minimum near the highest transmission. From the beam current data, the transmission efficiency is asymmetric, where scraping the same amount of input beam current leads to different output beam currents, depending on whether the positive or negative horizontal direction was scraped. The monotonic losses and transmission efficiency asymmetry suggest a horizontally asymmetric beam profile propagating through the DTL, which may originate from the 750 keV collimator elliptical aperture or non-Gaussian upstream effects.

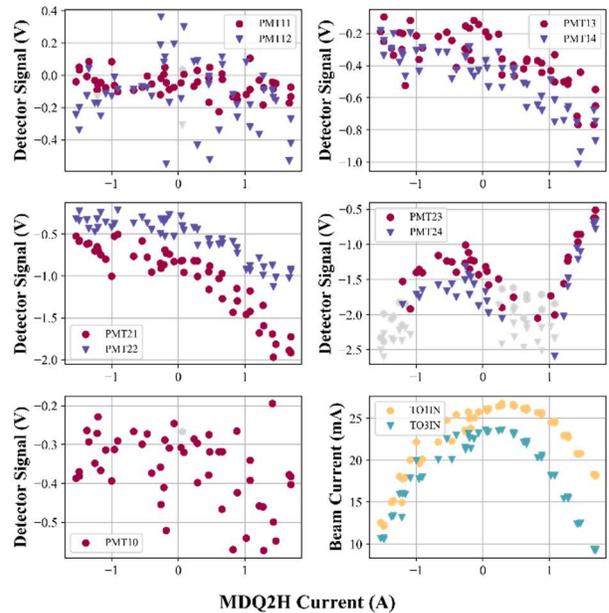

Figure 4: Response to MDQ2H. Gray points indicate saturation at the digitizer. PMT10 was parked at the most upstream position of the track. TO1IN and TO3IN are toroids at the entrance of DTL cavities 1 and 3, respectively.

To investigate the performance of the translating detector, which did not show significant changes in loss in response to beam scraping (Fig. 4), data were collected over the full length of the translating detector's track (Fig. 5).

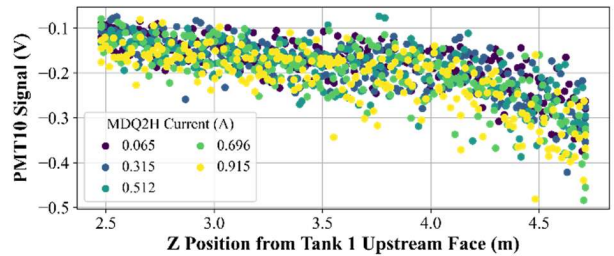

Figure 5: PMT10 loss signal along 2.2 meters of the first DTL cavity at different MDQ2H currents.

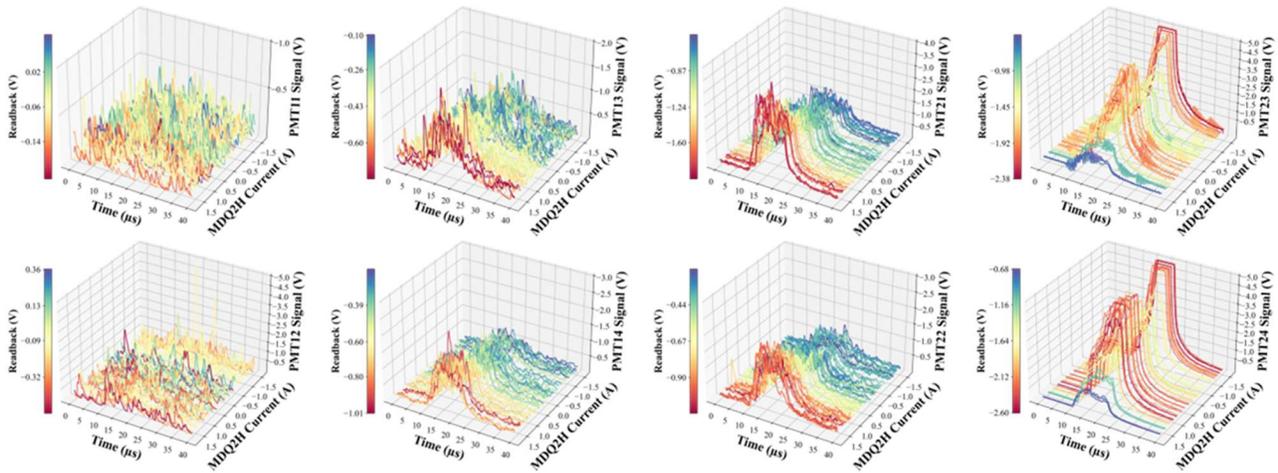

Figure 6: Response of the stationary PMTs to changes in horizontal dipole magnet, MDQ2H. Color is associated with the value of the digitized loss readback, plotted in Fig. 4.

Losses increase as the cart moves to a more downstream position, supporting the conclusion that the digitized detector signal is real. However, PMT10 is largely insensitive to the ~4 A changes in MDQ2H strength. We expect the low signal of PMT10 and stationary detectors PMT11 and PMT12 to be largely caused by the low energy, 750 keV, at the entrance to the DTL. Light output from the scintillators increases with the energy of the incoming particle [3], with gamma rays only detected above 5 MeV.

## ASSESSING ALIGNMENT

Given the drift tubes are offset up to 3.5 mm vertically, compared to 1.8 mm horizontally [1], we executed the same dipole study, but with a vertical dipole magnet, MDQ2V, at the same location upstream of the first DTL cavity (Fig. 7).

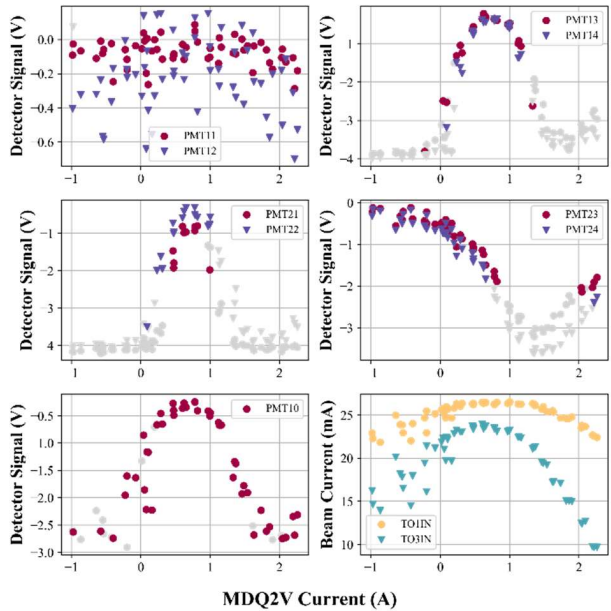

Figure 7: Response to MDQ2V.

The most upstream detectors (PMT11, PMT12) still yield no significant change in beam loss, however all other PMTs display much larger changes in beam loss throughout the 4 A change in MDQ2V vs. MDQ2H. Transmission is preserved at MDQ2V between approximately 0 and 1 A, and in this region, minima in some loss profiles are observed. The symmetry and local minima of beam losses suggest a more symmetric vertical profile than horizontal profile. The higher losses in the vertical plane may be connected to the more significant vertical offsets of the drift tubes, however, changes to MDQ2V more significantly affects beam centroid and envelope than MDQ2H and likely contribute to the increased losses (Fig. 8).

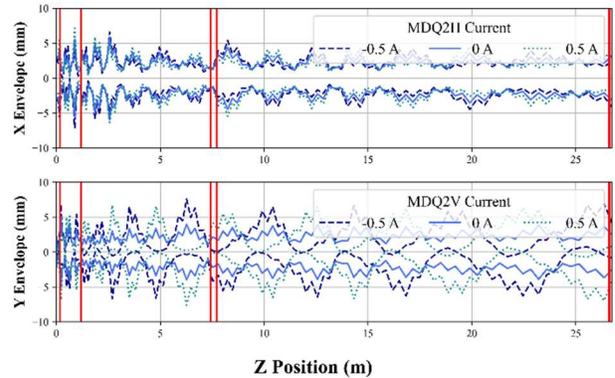

Figure 8: Simulated envelope response to changes in MDQ2H and MDQ2V. Red lines indicate PMT positions.

## CONCLUSION

We have implemented a significant upgrade to the scintillator-based loss detector system located at the first two DTL cavities in the Fermilab Linac, bringing the detectors to an operations-ready state. With the new system, we detected loss at beam energies greater than 10 MeV in response to small, tuning-relevant dipole kicks. Additionally, we've demonstrated an asymmetric horizontal loss response; however, this contrasts against the higher losses associated with vertical trajectory changes. With the system fully functional, we intend to utilize the system in collaboration with simulation results to assess sources of beam loss and improve tuning of the DTL.